\documentclass[twocolumn,
 aip,
 amsmath,amssymb,
reprint, %
]{revtex4-1}
\usepackage{amsmath}
\usepackage{graphicx}% Include figure files
\usepackage{dcolumn}% Align table columns on decimal point
\usepackage{bm}% bold math
\usepackage[bottom]{footmisc}
\usepackage{hyperref}
\usepackage[utf8]{inputenc}
\usepackage[T1]{fontenc}
\usepackage{mathptmx}
\usepackage{blindtext}
\usepackage{enumitem}
\usepackage{xcolor}
\usepackage{float}
\begin{document}

\preprint{AIP/123-QED}

\title{Nonlinear modeling and characterization of ultrasoft silicone elastomers}
% Force line breaks with \\

\author{Asimanshu Das}
 \affiliation{School of Engineering, Brown University, Providence, Rhode Island 02912, USA}%

%\author{Asimanshu Das}
%\author{Varghese Mathai\footnote{varghese\_mathai@brown.edu}}
%\author{Kenneth Breuer\footnote{kenneth\_breuer@brown.edu }}

\author{Kenneth S. Breuer}
\email{kenneth\_breuer@brown.edu }
\affiliation{School of Engineering, Brown University, Providence, Rhode Island 02912, USA}

\author{Varghese Mathai}%
\email{varghese\_mathai@brown.edu}
\affiliation{School of Engineering, Brown University, Providence, Rhode Island 02912, USA}%

\date{\today}% It is always \today, today,
             %  but any date may be explicitly specified

\begin{abstract}

We introduce a strain-energy based nonlinear hyper-elastic formulation to model the material properties of ultrasoft dielectric elastomers over a wide range of elastic properties, prestretch, and thicknesses. We build on the uniaxial Gent formulation, and under the conditions of equi-biaxial strain, derive an expression for the bulge deformation versus pressure. A circular bulge test methodology is developed to experimentally measure the mechanical response of the silicone membranes. The Gent model captures both neo-Hookean and strain-stiffening behaviors, and gives predictions which are in agreement with experimental measurements. Membranes with different thinner fractions are characterized over nearly one order of magnitude variation in shear modulus. Stiffer membranes are observed to harden at lower stretch ratios due to the increased fraction of polymer chains in them. The present approach offers a simple and cost-effective procedure for characterizing soft membranes under commonly encountered biaxial deformation conditions.
\end{abstract}

\maketitle
%\subsection*{Introduction}

Soft elastomeric materials find applications in a variety of fields spanning from tissue engineering and soft robotics to targeted cell-culture and self-sensing devices. 
%The need to replicate these motions in conventional robotics poses challenges and involves complexities. 
They form an essential ingredient in bio-inspired engineering today, owing to their resemblance to organic tissues, tunability of mechanical response, capacity for electrical actuation (dielectric), and stable and non-reactive properties\cite{yuk2016skin,stout2017electrostatically,roche2014bioinspired,Bohnker2019,song2008aeromechanics}. 
%Hauser et. al \cite{hauser2018compliant} developed a compliant universal gripper which is used for legged locomotion. They switch states of the soft membrane by changing the applied pressure and voltage across it which allows them to successfully maneuver in different terrains. Anderson et. al \cite{anderson2010thin} presented a membrane dielectric elastomer actuator (DEA) in which the phased actuation of electrode sectors of the motor imparts orbital motion, which in turn rotates a rotor. The ability of the system to develop high torque in a low mass body and to generate peak power at low rotational peak speeds makes it desirable to be used for robotic applications. 
%Song et. al \cite{song2008aeromechanics} have shown how thin, compliant membranes dictate the aerodynamic performance of flying mammals like bats. Their study confirms compliant wings lead to delayed stall at higher angles of attack and higher lift slope. 
%Furthermore, they provide the feasibility to tune material properties such as elastic modulus over wide ranges.  
While there are numerous potential applications of highly compliant soft materials, an essential step in their effective usage is to quantify their elastic response over a range of strains.  %characterization  is conventionally performed under uniaxial tensile test conditions\cite{sahu2014characterisation,goulbourne2011constitutive,malavc2005elastomers}.
%Uniaxial test has been often employed in material characterization because of its simplicity but the key issue here is the handling of the extremely soft, thin membranes during the sample preparation.
Material characterization studies have mostly focused on acrylic-based elastomers\cite{wissler2007mechanical,michel2010comparison,sahu2014characterisation,goulbourne2011constitutive,malavc2005elastomers}, which display significant viscoelastic losses\cite{palakodeti2006influence,Bohnker2019}. In comparison, silicone-based dielectric elastomers have comparatively lower losses\cite{Bohnker2019}, offer greater control of elasticity and are well-suited for in-house fabrication by combining a base mixture with a thinner component. They can attain very large strains before reaching yield point (hyperelasticity), with strain-stiffening occurring at large deformations\cite{hajiesmaili2019reconfigurable,leger2007adhesion,rashid2012mechanical}.

A number of hyper-elastic material models have been proposed in literature to model the non-linear response (stress vs. stretch-ratio) of soft materials.
%\cite{ogden1972large,arruda1993three,yeoh1993some,gent1996new}. 
Among these, the  Ogden\cite{ogden1972large}, Yeoh\cite{yeoh1993some} and Arruda-Boyce\cite{arruda1993three} models all sufficiently capture the material behavior through complex multi-parameter fits. However, besides the shear modulus parameter, the remaining fitting constants bear uncertain physical significance. In comparison, the hyper-elastic Gent model\cite{gent1996new} offers a simpler two-parameter constitutive relation composed of a shear modulus, $G$, and a locking parameter, $J_m$. Importantly, $J_m$ is representative of the strain-stiffening behavior, which can, in principle, be deduced from molecular considerations of the degree of polymer cross-linking\cite{hao2015hyperelasticity,beda2007modeling}. While the Gent formulation has a few inherent limitations due to its two parameter fitting\cite{pucci2002note}, it has been subject to several improvements in recent years\cite{destrade2017methodical,destrade2015extreme}.

Uniaxial tensile testing is the most widely used method for material characterization, owing to its simplicity and one-dimensional stretch condition. The testing is conducted in a Universal Testing Machine (UTM) and requires careful preparation of a thin sample (dog-bone shaped) in order to avoid stress concentrations near the gripped region. For soft materials, these can be challenging as sliding near the grips and thinning due to clamping stresses can affect the quality of results\cite{plante2006dielectric, hossain2012experimental}.  Furthermore, a number of assumptions are involved in translating the one-dimensional material response to commonly seen experimental conditions of plane or volumetric strains. A viable alternative is presented by the circular bulge method\cite{small1992analysis,huang2007mechanical}. Here, an applied differential pressure causes a thin material to enlarge under equi-biaxial stretch conditions. While traditionally employed for characterizing sheet metals and thin strips\cite{vlassak1992new,gutscher2004determination,ranta1979use}, recently, it has begun to be considered for soft materials such as hydrogels and PDMS\cite{sheng2017bulge,raayai2019volume}. The biaxial state of strain calls for a separate derivation of the stress vs. stretch-ratio relation.

In this Letter, we study the non-linear deformations of silicone elastomer membranes subjected to equi-biaxial state of stretch. A constitutive relation is derived based on a two-parameter hyper-elastic formulation~(\citet{gent1996new} model). We adopt a circular bulge test~(CBT) methodology, with pressure control, to obtain the deformation vs. pressure of membranes with different prestretch and elastic moduli. The model's predictions are compared with experiments, and a series of validations are conducted.  Following these, membranes with various thinner fractions are characterized, and their hyper-elastic parameters (shear modulus and locking parameter) are estimated.

The  membranes were fabricated from a platinum-based addition-cure type silicone rubber (Dragon Skin FX Pro-Shore Hardness 2A, Smooth-On Inc., Macungie, PA). The base material was composed of two constituent components: Part A, a silicone hydride and Part B, a vinyl compound that acts as a catalyst. To this mixture, a thinner component (TC 5005-C, BJB Enterprises Inc., Tustin, California) was added to control the material properties of the cured elastomer~(see supplemental material for details of the fabrication).  The membranes thickness, $h$, was measured using a microscope stage (See supplemental material).  Thicknesses ranged from 200 -- 1000 $\mu$m, which is small compared to the lateral dimension, $D$, of the membrane disk ($h/D \leq 0.01$). Therefore, the bending stiffness is assumed to be negligible when compared to the tension in the membrane. 

\begin{figure}[!t]
	\centering{
		\includegraphics[width=0.45\textwidth]{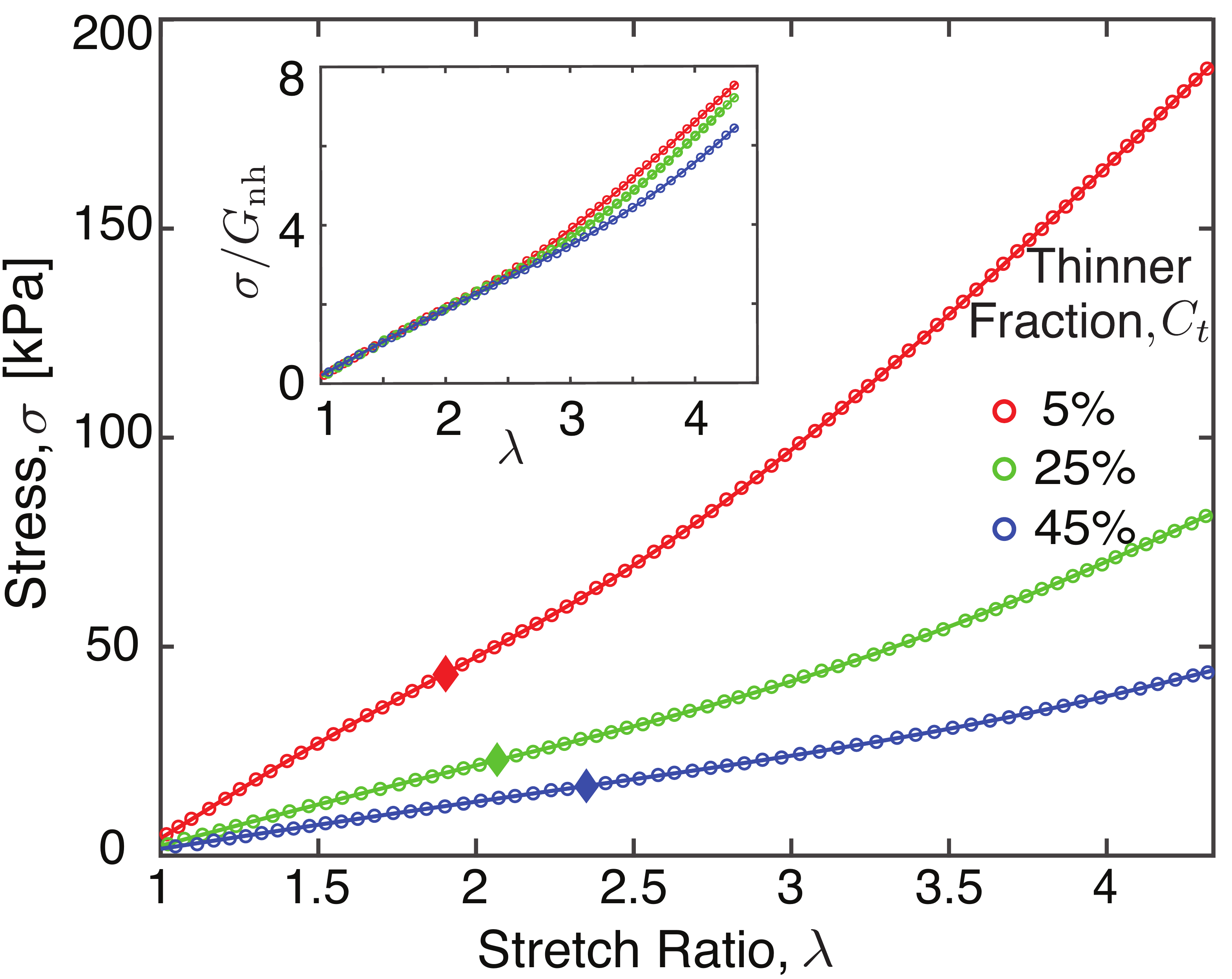}}
	\caption{Uniaxial stress vs. stretch-ratio data (circles) for silicone membrane samples with thinner fractions, $C_t$ = 5\%, 25\% and 50\%. The solid curves show the Gent model fit to the data. The diamond symbols denote the respective onsets of strain stiffening in the samples, calculated from the inflection points of the respective curves. The inset presents stress divided by the best-fit neo-Hookean shear modulus, $G_\text{nh}$, which portrays clear evidence of the varying degree of strain-stiffening for the three cases.}
	\label{fig:stretch-stress}
\end{figure}

For uniaxial testing, dog-bone shaped samples were cut from membranes using a laser cutter and gripped between the jaws of a UTM~(Instron 5942). Fig.~\ref{fig:stretch-stress} shows the stress vs. stretch-ratio curves for three representative cases with thinner fractions 5\%, 25\% and 50\% by weight.

The curves do not follow a linear elastic behavior, and the material displays strain-stiffening for large values of $\lambda$, the onset of which is qualitatively marked by the diamond symbols in Fig.~\ref{fig:stretch-stress}. The inset in the figure shows the same data normalized using a neo-Hookean fit: $\sigma = G_\text{nh}(\lambda - 1/\lambda^2)$, where $G_\text{nh}$ are the respective shear moduli obtained using a least-squares fitting. For low stretch-ratios, the single-parameter neo-Hookean fit captures the material response well, while the model deviates significantly at large deformations. Interestingly, the point of departure from neo-Hookean behavior is a function of the thinner fraction, with the onset of strain-stiffening occurring at larger stretch-ratios for the softer materials. 

To better describe the nonlinear behavior, we introduce the two-parameter Gent model\cite{gent1996new}, which, for uniaxial stretch condition, is written as
\begin{equation}
\sigma=G_m (\lambda-\dfrac{1}{\lambda^{2}}).
\label{Gent_uniax}
\end{equation}
Here, $G_m \equiv {G J_{m}}/{(J_{m}-I_{1}+3})$, where $G$ is the shear modulus, $J_m$  is the so-called ``locking parameter" and $I_1$ is the first invariant of the left Cauchy-Green deformation gradient tensor\cite{bower2009applied}. The Gent relation (Eq.~\ref{Gent_uniax}) was fit to the uniaxial experimental data, which yielded the material constants, $G$ and $J_m$. These constants will be used to evaluate the accuracy of the bulge test predictions.

\begin{figure}[!bp]
	\centering{
		\includegraphics[width=0.45\textwidth]{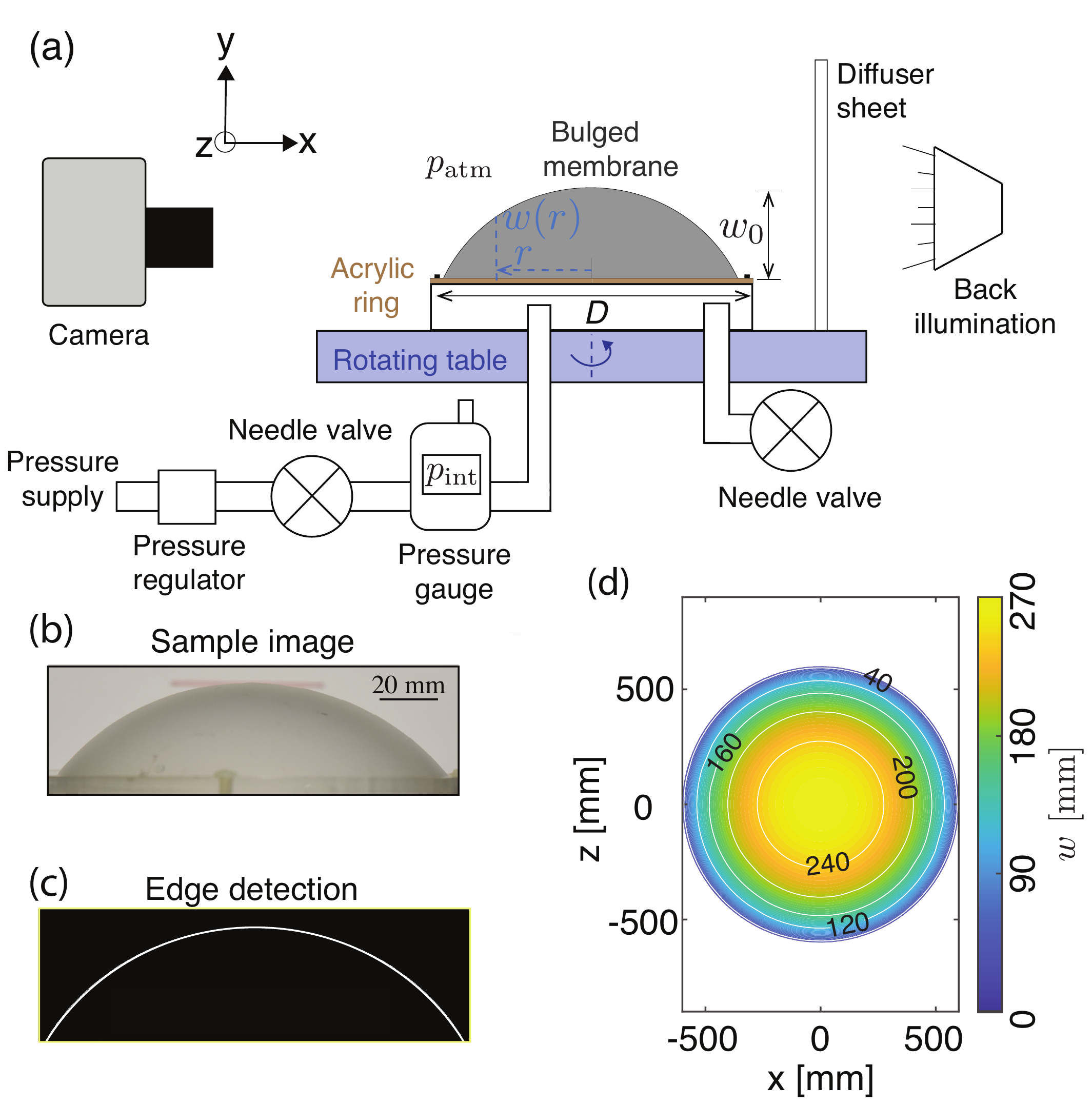}}
	\caption{Circular bulge testing procedures. (a) Schematic illustration of the bulge experimental setup. The needle valve regulates the pressure from the high-pressure supply. The bulge chamber is placed on top of a rotating table in order to capture the bulge from different angular orientations. (b) Sample image captured at an arbitrary deformation. (c) Edge detection on the sample image in (b). (d) Top view of the surface plot acquired by combining the deformation profiles from different angular views.}
	\label{fig:Bulge_setup}
\end{figure}

For bulge testing, a circular cut-out (10 cm in diameter)  was laser cut from a cured sheet of membrane and the membrane transferred onto a custom-built bi-axial stretching apparatus (see supplemental material Fig.~S-3), which was used to achieve the desired pre-stretch, $\lambda_0$. A circular acrylic ring of inner and outer diameters 120 mm and 127 mm, respectively, was glued onto the stretched membrane using a silicone adhesive (Silpoxy, Smooth-On Inc., Macungie, PA).  Once the adhesive had cured~($\sim$ 60 mins), the membrane disk was mounted onto the bulge testing apparatus (Fig.~\ref{fig:Bulge_setup}a) using a rubber gasket to ensure that the  chamber was air-tight. 

%
%% This belongs in the Supp Matt% A silicone adhesive tape was applied onto each finger of the stretching apparatus, which ensured proper adhesion. Graduations on the bi-axial stretcher allowed us to impart the desired pre-stretch\cite{rosset2016fabrication}. The pre-stretched membrane was then glued onto a thin circular acrylic ring with a silicone glue (Smooth-On Silpoxy,).
%
% Figure~\ref{fig1}(a) shows a schematic of  a bulge test setup used testing the membranes. An initially flat circular membrane with a diameter $D$  is fixed onto the setup, and a uniform pressure $p$ was regulated using using needle valve and high pressure supply. The membrane deforms into a nearly hemispherical shape~(see Fig.~\ref{fig1}(b)). The profile of the bulged membrane was detected  of theHere to develop the theoretical model, we assume an incompressible hyper-elastic material, with thickness $h$  and with a pre-stretch $\lambda_0$.
A pressure regulator (Bellofram Type 70-Range 0-2 PSIG, Marsh Bellofram Corporation, Newell, WV), in combination with a bleed valve (Swagelok Stainless Steel Low Flow Metering Valve, Swagelok Company, Solon, OH), was used to establish the pressure in the test chamber, $p = p_\text{int} - p_\text{atm}$, which was monitored and recorded using a differential manometer~(Reed R3001) {\color{black} with a resolution of $10^{-3}$ kPa}. Images of the bulging membrane were recorded using a digital camera (Nikon D7200, 24.2 megapixel resolution), illuminated with a uniform back-light to ensure good image contrast~(Fig.~\ref{fig:Bulge_setup}b). \textcolor{black}{A checkerboard pattern, with a grid size of 10~mm, was used for calibration. The resolution of the image was 12 pixels/mm.} The {\it Canny} edge detection algorithm was employed to extract the membrane shape (Fig.~\ref{fig:Bulge_setup}c). The apparatus was mounted on a rotating table so that images could be acquired from multiple positions in order to reconstruct the deformation of the entire membrane surface, $w(r,\theta)$, (Fig.~\ref{fig:Bulge_setup}d) and the centerline deformation, $w_0$. 

Data was taken over a range of pressures until the normalized bulge deformation, $w_0/D$, reached approximately  $0.5$.  The membranes were observed to bulge nearly axisymmetrically (Fig.~\ref{fig:Bulge_setup}d), with $w(r)$ obeying a spherical cap deformation. Assuming a uniform applied pressure, $p$, and a spherical cap profile for the bulged membrane, the effective stretch-ratio in the deformed state can be approximated as \cite{waldman2017camber}
%Under the assumption of a spherical cap profile\cite{waldman2017camber}, the membrane curvature can be estimated from the measured centerline deformation, $w_0$, using the relation $\kappa={8w_{0}}/({D^2+4w_0^2})$. With the curvature known, the stretch-ratio $\lambda$ in the deformed state can be approximated as:
\begin{equation}
\lambda= \dfrac{2\lambda_{0}}{\kappa^{*}}\sin^{-1}({\kappa^{*}}/{2}),
\label{lambda_kappa_eq}
\end{equation} 
where $\kappa^{*} = \kappa \ D$ is the dimensionless curvature, and $\lambda_0$ is the pre-stretch ratio~(see supplemental material for details).

%\textcolor{black}{In order to accurately model the deformation observed in the pressure bulge relation, a simple neo-Hookean formulation is not adequate as in Fig.~\ref{fig:stretch-stress}, so we need to take into account the strain-stiffening behavior of such hyperelastic materials.}
%
%The Gent hyper-elastic model\cite{gent1996new} is able to account for both neo-Hookean and strain-stiffening behaviors. For uniaxial-stress condition, it is given by
%\begin{equation}
%\sigma=G_m (\lambda-\dfrac{1}{\lambda^{2}}),
%\label{Gent_uniax}
%\end{equation}
%
%where $\lambda$ is  the stretch ratio and $G_m \equiv {G J_{m}}/{(J_{m}-I_{1}+3})$ is an effective shear modulus. Here,  $G$ is the shear modulus, $J_m$ is a so-called ``locking parameter", and $I_1$ is the first variant of the left Cauchy-Green deformation gradient tensor\cite{bower2009applied}.

\begin{figure}[!tbp]
	\centering{
		\includegraphics[width=0.45\textwidth]{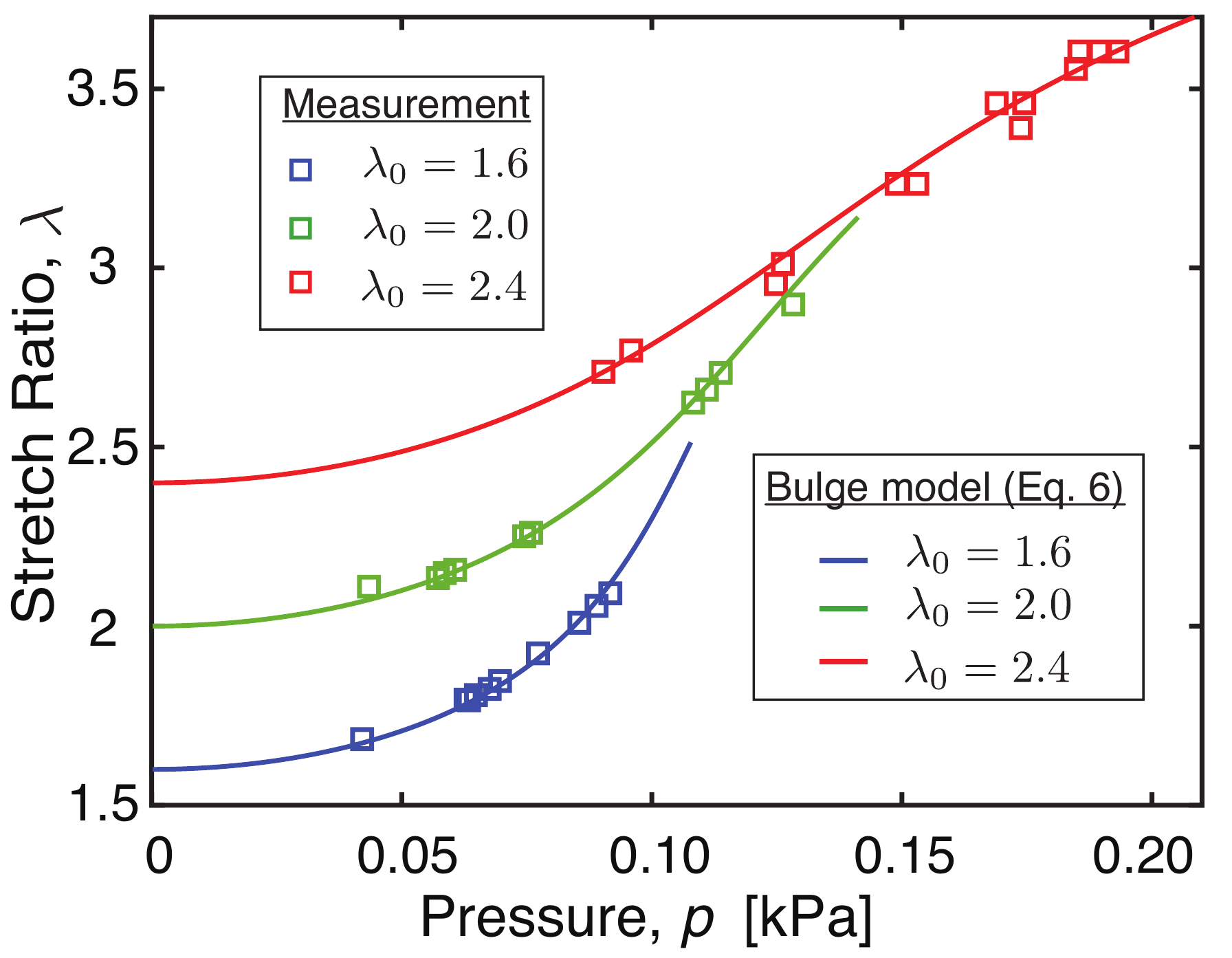}}
	\caption{Stretch-ratio vs. bulge pressure from experiments~(square symbols) and the Gent biaxial model predictions (curves) for membranes with thinner fraction, $C_t = 25\%$, for  three different pre-stretches, $\lambda_0 =$ 1.6, 2 and 2.4. In all of the three cases shown, the membrane deforms from a flat circular to a hemispherical shape, which corresponds to the stretch-ratio increase from $\lambda_0$ to $\frac{\pi}{2} \lambda_0$. \textcolor{black}{ The errorbars~(estimated from three independent measurements) are smaller than the symbol size}.}
	\label{fig:pressure-strech ratio}
\end{figure} 

The uniaxial Gent relation, used above, can be generalized to the axisymmetric bulge condition. During the bulge process, the membrane expands uniformly, and so the state of deformation can be approximated as equi-biaxial strain. Assuming an isotropic incompressible material, the deformation gradient tensor ($\mathbf{F}$) can be written as a diagonal matrix with  elements: $\lambda, \lambda$, and $\dfrac{1}{\lambda^{2}}$. The Cauchy stress ($\mathbf{\sigma}_b$) for incompressible Gent model can be written as
\begin{equation}
\mathbf{\sigma}_b=-p\mathbf{I}+G_m\mathbf{B}, 
\end{equation}
where {\bf I} is the identity matrix and $\mathbf{B}=\mathbf{F}.\mathbf{F^{T}}$ is the Cauchy-Green tensor. So the in-plane stresses are,
\begin{equation}
\sigma_{11}=\sigma_{22}=G_m(\lambda^{2}-\dfrac{1}{\lambda^{4}}).
\end{equation}

The resulting tension in the bulged membrane, $T_b$, is derived from the  first Piola-Kirchhoff stress as $T_b=G_m h(1-{1}/{\lambda^{6}})$. With the tension, $T_b$, in the bulged membrane known, the static equilibrium shape under uniform pressure loading can be obtained from the Young–Laplace equation \cite{waldman2017camber},
\begin{equation}
\kappa+ \dfrac{p}{T_b}=0.
\end{equation}
Substituting the constitutive relations~(see supplemental material), we obtain a relation between the pressure and stretch ratio as
\begin{equation}
p \approx\dfrac{10G_mh}{D}(1-\dfrac{1}{\lambda^{6}})\sqrt{{\lambda}/{\lambda_{0}}-1}
\label{eq_press}
\end{equation}

%
%In order to account for the stiffening behaviour at large deformations, we had to take up a constitutive material model which effectively captures the material behaviour at all deformation states. The neo-Hookean model is widely regarded due to its simple mathematical formulations, so we took up the Gent model which reduces to neo-Hookean for limiting value of the first variant of the left Cauchy-Green deformation gradient tensor.

Experimental measurements from the circular bulge test (CBT) of the  stretch ratio vs. pressure are indicated by the square symbols in Fig.~\ref{fig:pressure-strech ratio}.  The solid lines show the Gent model predictions (Eq.~\ref{eq_press}), where the constants, $G$ and $J_m$, are determined from previously conducted uniaxial tests. Here, the blue, green and red curves correspond to three different pre-stretches, $\lambda_0 =$ 1.6, 2.0 and 2.4, respectively and all three show excellent agreement between measurements and theory.  Interestingly, for the lowest value of pre-stretch, $\lambda_0=1.6$, the material response lies in the neo-Hookean regime, whereas strain-stiffening is noticeable for the highest pre-stretch case, $\lambda_0=2.4$. Similarly, the predictions match for independent variations in pre-stretch, $\lambda_0$, and membrane thickness, $h$, (Fig.~S-5, supplemental material). These validations establish the robustness of the two-parameter Gent formulation. Note that, at sufficiently large stretch-ratios, Eq.~\ref{eq_press} reduces to $p \approx \frac{10G_m h}{D}\sqrt{{\lambda}/{\lambda_0} -1}$. For the largest stretch-ratio  attained in the present work~($\lambda \approx 3.5$ in Fig.~\ref{fig:pressure-strech ratio}), this approximation is reasonable. The current model is applicable up to a maximum deformation, $w_0/D = 0.5$, or, equivalently, ${\lambda}/{\lambda_0} \leq \pi/2$. Beyond this limit, the spherical cap assumption for the deformation is no longer valid.

%{\color{red} For discussion section: It was noted that even slight non-uniformities in thickness or pre-stretch led to non-axisymmetric deformation, thus leading to erroneous measurements. Therefore, the making of reliable and reproducible circular membrane rings is crucial for the accuracy of the bulge test.}

\begin{figure} [!tp]
	\centering{
		\includegraphics[width=0.48\textwidth]{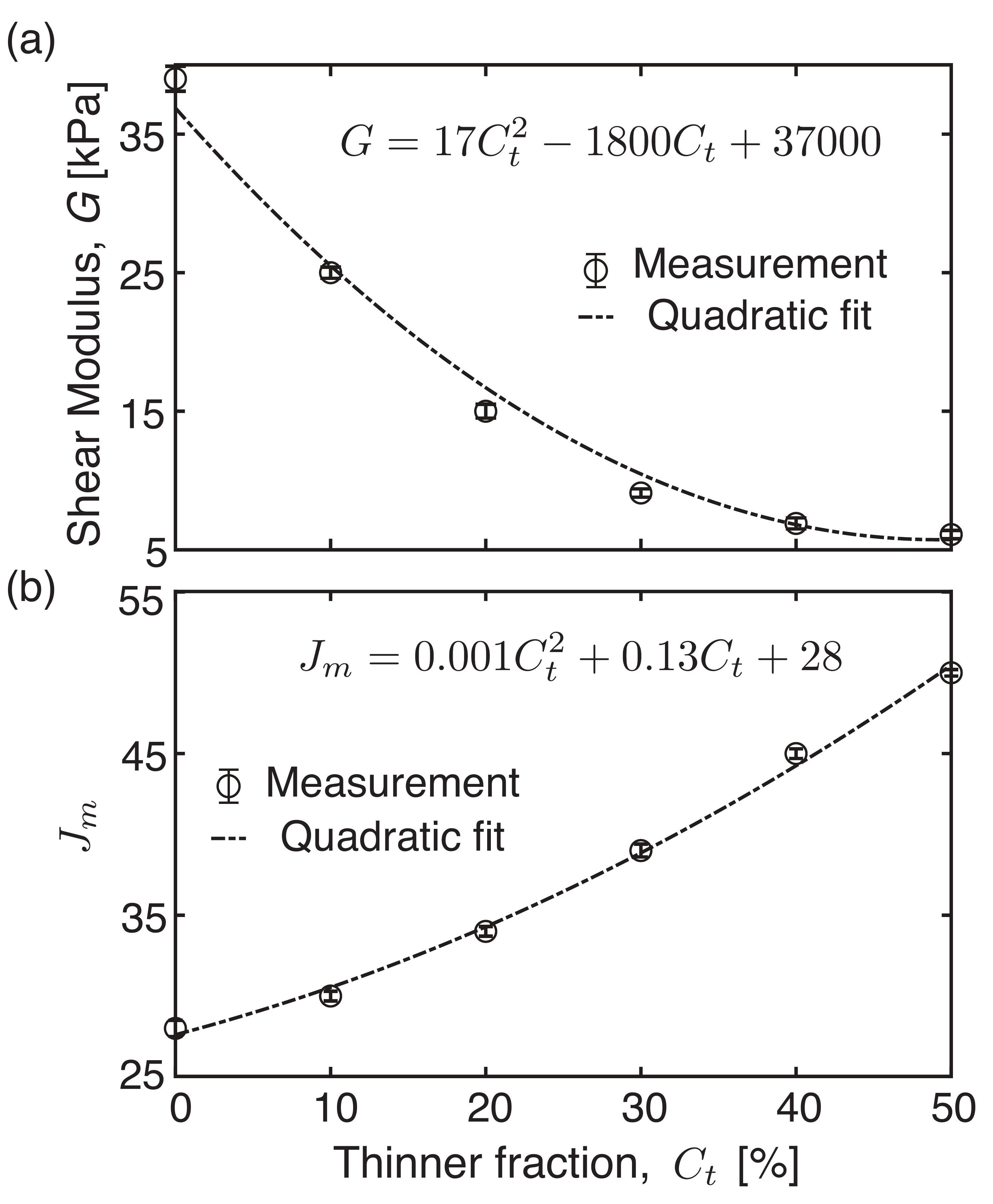}}
	\caption {Gent hyper-elastic constants obtained using the circular bulge testing methodology. (a) Shear modulus ($G$) for varying weight percentages of thinner, $C_t$. (b) Dependence of the locking parameter ( $J_{m}$ ) on the thinner percentage. $J_m$ is a representative of the onset of stiffening. Error bars represent maximum experimental variation from the mean. The polynomial fits in (a) and (b) provide useful relations to obtain the material properties for arbitrary choice of thinner percentage $C_t$.}
	\label{fig:Hyperelastic-parameters}
\end{figure}

Having demonstrated that the predictions of the Gent model are in good agreement with experimental measurements over a range of pressures, we now characterize the modulus, $G$, and the locking parameter, $J_m$, for membranes materials with varying thinner concentration, $C_t$ (Fig.~\ref{fig:Hyperelastic-parameters}a, b).  Overall, by varying $C_t$ in the range: 0 -- 50 \%,  the material's stiffness, $G$,  can be modified by nearly an order of magnitude, which demonstrates the wide tunability in properties achievable with these ultra-soft silicone elastomers. Concurrently, the locking parameter, $J_m$, increases with $C_t$, indicating a delay in the  onset of strain-stiffening. This can be explained by the reducing fraction of cross-linking polymers. Thus, with growing thinner concentration, the membrane behavior tends toward the simpler neo-Hookean solid\cite{horgan2015remarkable}, which is nicely captured in the Gent formulation.
%The phenomenological form of the Gent model enables it to approach the classical neo-Hookean model  as the locking parameter, $J_m$, tends to a large value\cite{horgan2015remarkable}.}

%Least-square fits to the experimental data are shown, which provides useful relations for estimating the hyper-elastic parameters at intermediate thinner fractions.
%In principle, the shear modulus ($G$) and the locking parameter ($J_m$) can be quantified on the basis of the cross-linking density\cite{gent1996new}. So the varying the thinner fractions results in altering the cross-linking thus, dictating the material's hyper-elastic characteristics.  

In summary, we have introduced a circular bulge test methodology for estimating the hyper-elastic parameters of highly compliant membranes for a range of membrane elastic properties, pre-stretches and thicknesses. A two-parameter phenomenological model (Gent model\cite{gent1996new}), based on the limiting chain extensibility, was adopted and extended theoretically for equi-biaxial condition to derive a relation between pressure and bulge deformation. The model was validated using independent measurements from uniaxial and circular bulge  tests, and following this, a systematic characterization of the membrane properties with changing thinner concentration was performed. Both the shear modulus, $G$, and the locking parameter, $J_m$, exhibit a dependence on the amount of thinner added. 
Since the locking parameter quantifies the onset of strain-stiffening, the current model can, in principle, be applied to a wider class of soft polymers, including hydrogels, organic elastomers and soft dielectrics. The theoretical formulation introduced here can also be extended to anisotropic membranes and to non-equiaxial stretch conditions. The methods introduced here provide a reliable means to characterize ultra-soft membranes, which can replace expensive testing procedures requiring Universal Testing Machines.
\vspace{-0.5 cm}
\subsection*{Supplementary material}\vspace{-0.3 cm}
	Supplementary material contains details of the biaxial modeling, the circular bulge test experiments, and videos of the stretching and unwrinkling procedure.\\
	
	We thank Anupam Pandey, Pradeep Guduru and Jillian~Bohnker for useful insights.  This research was supported by the US Army Natick Soldier Systems Center. The data that support the findings of this study are available from the corresponding author upon reasonable request.

%\bibliographystyle{prsty_allauthors}
%\bibliography{aipsamp}

%merlin.mbs aipnum4-1.bst 2010-07-25 4.21a (PWD, AO, DPC) hacked
%Control: key (0)
%Control: author (8) initials jnrlst
%Control: editor formatted (1) identically to author
%Control: production of article title (0) allowed
%Control: page (1) range
%Control: year (1) truncated
%Control: production of eprint (0) enabled
\providecommand{\noopsort}[1]{}\providecommand{\singleletter}[1]{#1}%

\clearpage
\newpage

%% Supplemental materials
\clearpage
\newpage
\setcounter{equation}{0}
\setcounter{section}{0}
\setcounter{figure}{0}
\setcounter{table}{0}
\setcounter{page}{1}
\makeatletter
\renewcommand{\theequation}{S-\arabic{equation}}
\renewcommand{\thefigure}{S-\arabic{figure}}
\renewcommand{\thesection}{S-\Roman{section}}
\renewcommand{\thesubsection}{S-\Roman{section}-\alph{subsection}}
\renewcommand{\thetable}{S-\arabic{table}}
\end{document}

\preprint{AIP/123-QED}

\title{~~~~~~~~~~~~~~~~~~~~~~~~~~Supplemental  Material \\ Nonlinear modeling and characterization of ultrasoft silicone elastomers}
% Force line breaks with \\

\author{Asimanshu Das}
 \affiliation{School of Engineering, Brown University, Providence, Rhode Island 02912, USA}%

%\author{Asimanshu Das}
%\author{Varghese Mathai\footnote{varghese\_mathai@brown.edu}}
%\author{Kenneth Breuer\footnote{kenneth\_breuer@brown.edu }}

\author{Kenneth S. Breuer}
\email{kenneth\_breuer@brown.edu }
\affiliation{School of Engineering, Brown University, Providence, Rhode Island 02912, USA}

\author{Varghese Mathai}%
\email{varghese\_mathai@brown.edu}
\affiliation{School of Engineering, Brown University, Providence, Rhode Island 02912, USA}%

\date{\today}% It is always \today, today,
             %  but any date may be explicitly specified

\maketitle
%\subsection*{Introduction}

\section{Membrane fabrication and testing }

The membranes were fabricated by curing a platinum-based addition-cure type silicone rubber (Dragon Skin FX Pro-Shore Hardness 2A, Smooth-On Inc., Macungie, PA). The product consists of two parts: Part A, a silicone hydride and part B, a vinyl compound that acts as a catalyst. To this mixture, a thinner component (TC 5005-C, BJB Enterprises Inc., Tustin, California) was added to control the material properties of the cured rubber. The three components were mixed together and poured onto a flat polyethylene terephthalate (PET) film of dimensions 200 mm $\times$ 300 mm and thickness 125 $\mu$m. \textcolor{black}{The PET film was placed on a leveled steel plate~(CNC milled with $1~\mu$m surface roughness).} A uniform wet-film thickness was obtained using a height-adjustable film applicator (TQC-SH0342-300, TQC Sheen B.V, LL Capelle aan den IJssel, Netherlands) {\color{black}, with a least count of 10 micron.} \textcolor{black}{As the membrane mixture has a low viscosity, it flows and equilibrates in thickness inside a dam-like structure.} 

\subsection{Membrane thickness characterization}

\begin{figure}[!htbp]
	\centering{
		\includegraphics[width=0.45\textwidth]{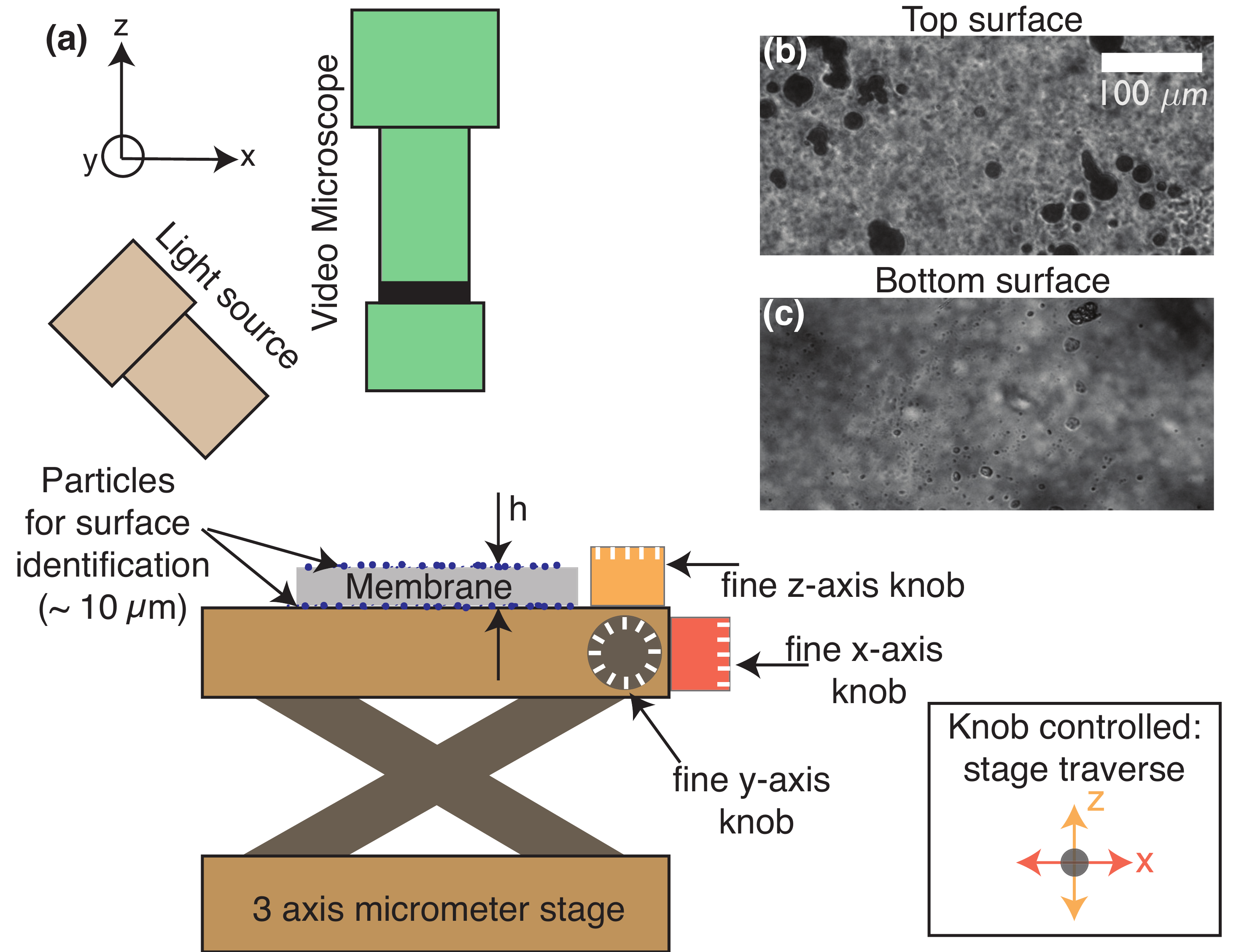}}
	\caption{(a) Schematic of the thickness measurement setup. The stage facilitates 3-axis movement with its micrometer positioning system (positioning accuracy of 1 $\mu$m). Here, the z-axis knob  controls the vertical movement of the stage. Talcum powder particles are sprayed to ensure better identification of the top and bottom surfaces. (b) \& (c) show images of the top and bottom surfaces of the membrane, respectively. The darkened spots are the talcum powder particles.}
	\label{figS:Microscope_thickness}
\end{figure}

The fabricated membranes were extremely thin and compliant. A non-intrusive method, comprising a video microscope and a 3-axis stage, was employed. The height of the stage was precisely controlled with a micrometer positioning system (positioning accuracy up to 1 $\mu$m), that provided  a robust method to not only measure thickness at a point, but also scan the entire surface to quantify for the variation in thickness. Here, the z-axis is the vertical movement of the stage. Talcum powder particles~($\sim$ 10 $\mu$m particles) were sprayed to ensure better identification of the top and bottom surfaces, as shown by the darkened spots in Fig.~\ref{figS:Microscope_thickness}b, c. {\color{black}The membranes, upon drying, became thinner than the original wet-films. The dry-film thicknesses of the fabricated membranes lies in the range: 300 $\mu$m to 1500 $\mu$m, with a thickness precision of $\pm$ 50~$\mu$m. For any chosen membrane, the thickness non-uniformity along the surface was $\pm$ 7.5 $\mu$m.}

\subsection{Membrane release}
\begin{figure}[!b]
	\centering{
		\includegraphics[width=0.45\textwidth]{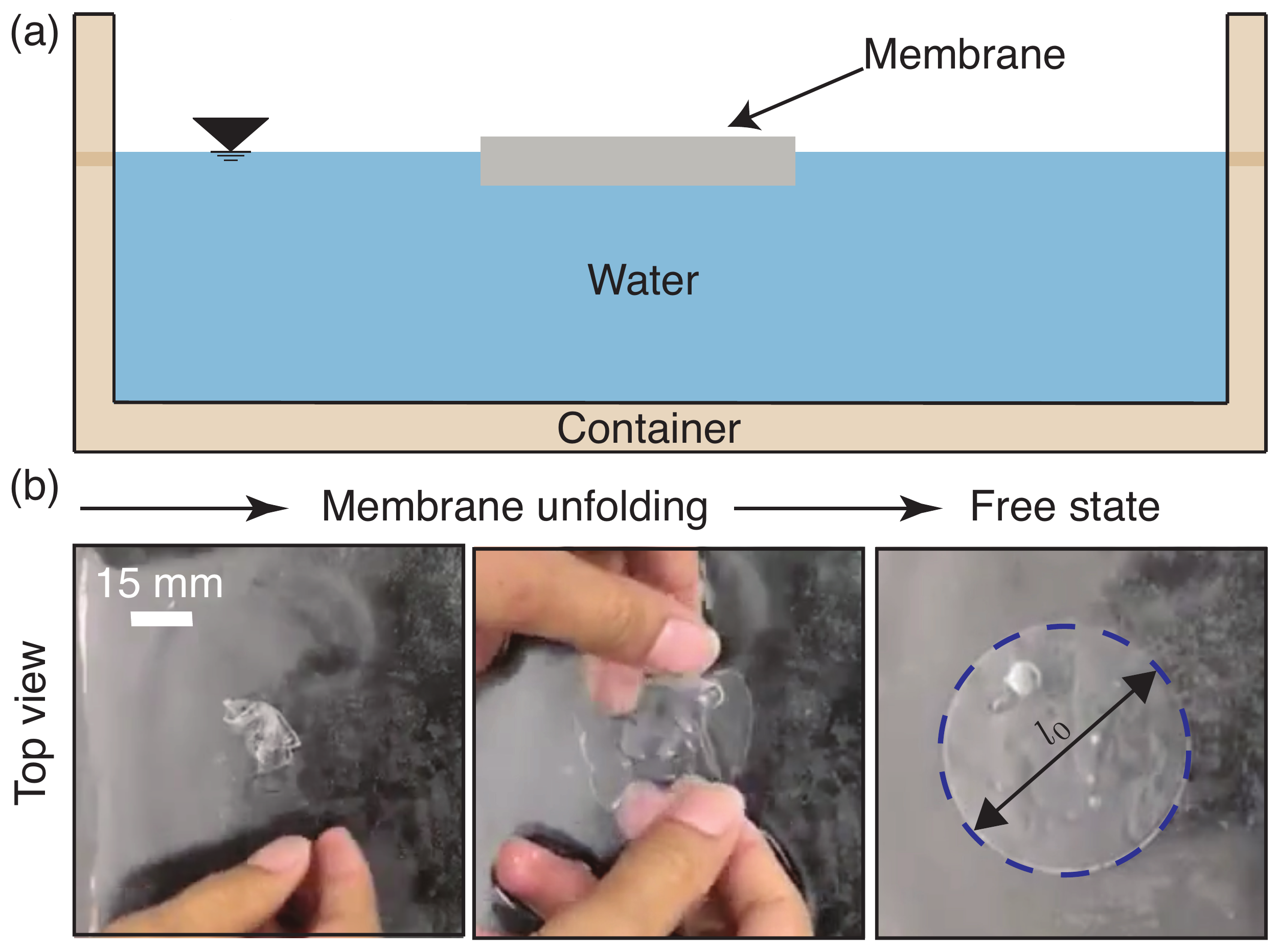}}
	\caption{(a) Depicts a floating membrane on a water bath immediately after release from the PET film. (b) It shows the procedure of unwrinkling  the  membrane. The circular membrane has relaxed to it initial diameter (right side image), denoted by $l_0$.}
	\label{figS:water_bath}
\end{figure}

Because the membranes are very thin and soft, they are difficult to handle, and tend to fold and wrinkle in the presence of residual electrostatic charges (due to peeling from the PET film). To unwrinkle the membrane, we suspended them in a water bath (as shown in Fig.~\ref{figS:water_bath}a). Since the membranes were nearly neutrally buoyant, this established  a state where they could relax to their true shape. Fig.~\ref{figS:water_bath}b shows the unwrinkling of the membrane to its free state when it had low residual stresses~(see also supplemental video S1). Note that immersion in the bath did not lead to any noticeable water absorption. 

\subsection{Uniaxial  testing}
Once cured, a ``dog-bone" shape measuring 38 mm gauge length and 12 mm width was cut out from the cured membrane film using a laser cutter.  The dog-bone sample was necessary to ensure a nearly uniaxial tensile load on the membrane. The sample was mounted onto  two pairs of metal plates using adhesive tape (4910 VHB, 3M Company, Saint Paul, MN) and the specimen was  gripped between the jaws of a Universal Testing Machine (Instron 5942). Load tests were carried out at a strain rate, $\dot \epsilon$, of 0.08 mm/s. This strain rate was chosen after carrying out tests at several values of $\dot \epsilon$ between 0.04~mm/s and 5~mm/s. Notably, the stress-strain curves showed little dependence ($\pm 2\%$) on the strain rate, $\dot \epsilon$. 
%Representative stress vs. stretch-ratio curves obtained from the uniaxial tests are shown in Fig.~\ref{fig:stretch-stress} of the main paper.

\subsection{Sample preparation for Biaxial stretching}
\begin{figure}[!t]
	\centering{
		\includegraphics[width=0.5\textwidth]{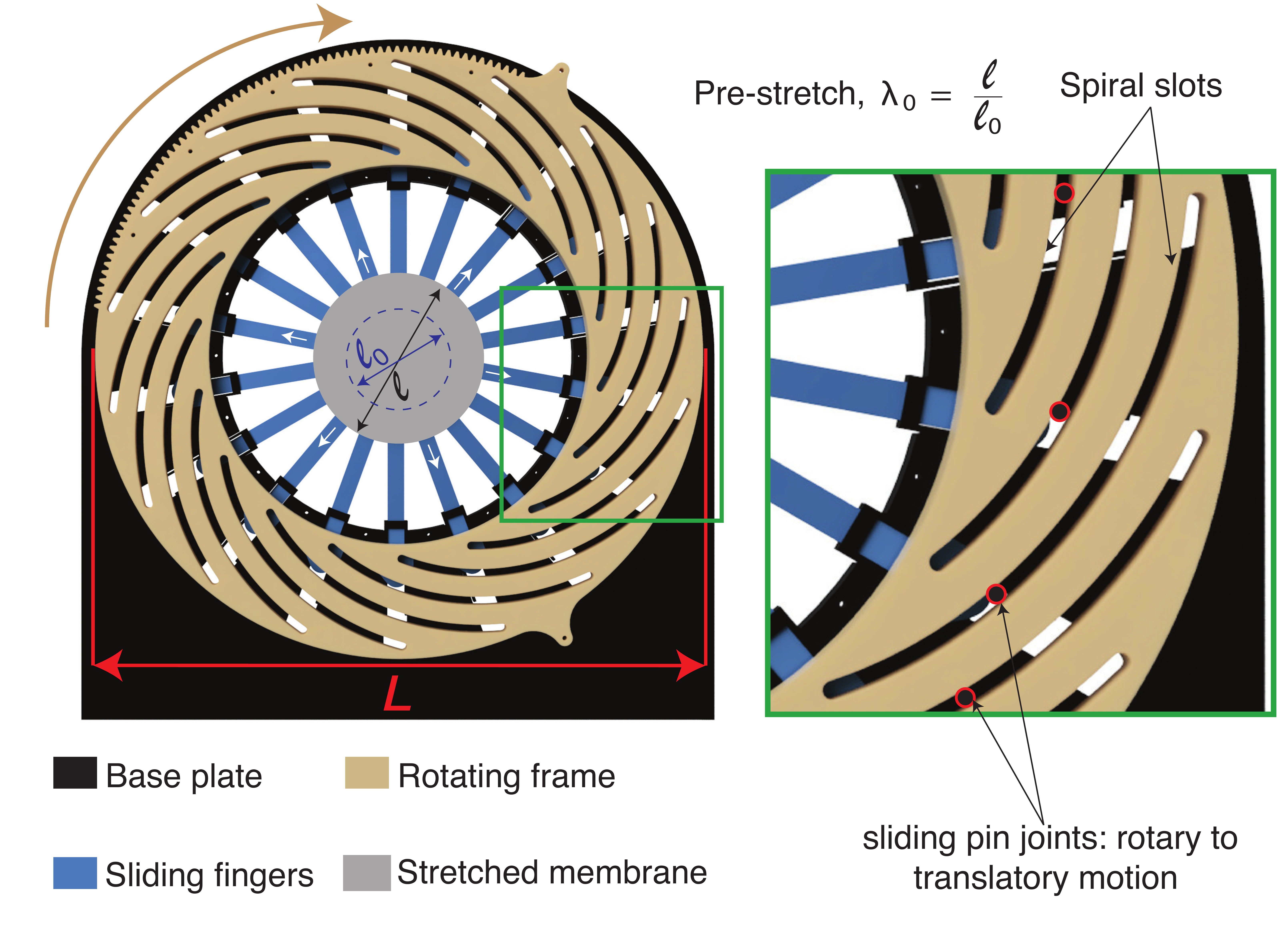}}
	\caption{Schematic of the top view of biaxial stretching apparatus. The sliding pin joints trace  the spiral profile, effectively converting the rotary  motion into radial motion of the sliding fingers. This enables a nearly equi-biaxial stretching. }
	\label{figS:Biaxial Stretcher}
\end{figure}

In order to provide an initial pre-stretch to the membrane, a  biaxial stretcher was designed, fabricated and used (Fig.~\ref{figS:Biaxial Stretcher}). The stretcher was designed in a way that rotation of the outer ring produced radial motion to the fingers holding the membrane sample. The stretcher is made from acrylic, and is composed of three main parts: (i) a base plate, (ii) a rotating frame and (iii) eighteen sliding fingers. The base plate houses the finger assembly and supports the rotating frame. The fingers distribute the stretching forces and ensure nearly equi-biaxial stretching. A representative example of the stretching operation  is shown in supplemental video S2. The stretcher has a minimum diameter of 70 mm, and can stretch up to 240~mm, which gives a wide range of pre-stretches $\lambda_0 \in [1.0, 3.4]$. The part files provided as supplemental material can be used to reproduce the design.

\section{Theoretical modeling}
The uniaxial stress ($\sigma$) for the Gent hyper-elastic model can be written as\cite{gent1996new},

\begin{equation}
\sigma=G_m(\lambda-\dfrac{1}{\lambda^{2}}),
\end{equation}
where $G_m=(\dfrac{G J_{m}}{J_{m}-I_{1}+3})$,  $G$ is the shear modulus, $J_m$ is the "locking parameter", and $I_1$ is the first invariant of the left Cauchy-Green deformation gradient tensor. The principal stretches for uniaxial extension are represented by $\lambda_1, \lambda_2$ and, $\lambda_3$. Here $\lambda_1=\lambda$ and $\lambda_2=\lambda_3$. Assuming the material to be incompressible,  $\lambda_1\lambda_2\lambda_3=1$. Consequently, the first invariant, $I_{1}=\lambda_{1}^{2}+\lambda_{2}^{2}+\lambda_{3}^{2}$, reduces to $\lambda^{2}+\dfrac{2}{\lambda}$.\\

In the bulge test, an initially flat circular membrane with a diameter, $D$,  evolves into a nearly hemispherical shape when subjected to uniform pressure, $p$. We assume an incompressible hyper-elastic material with thickness, $h$,  and pre-stretch, $\lambda_0$. Assuming a spherical cap profile, we can express the curvature, $\kappa$, of the membrane as,

\begin{equation}
\kappa=\dfrac{8w_{0}}{D^2+4w_0^2},
\end{equation}
where $w_0$ is the maximum deformation at the centre.

\begin{figure}[!h]
	\centering{
		\includegraphics[width=0.4\textwidth]{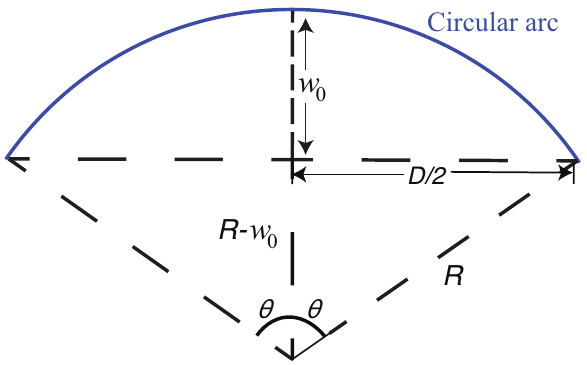}}
	\caption{Schematic of the cross-sectional view of the membrane deformation.}
	\label{figS:geometry}
\end{figure}

An illustration of the deformed segment is given by Fig.~\ref{figS:geometry}. Based on simple geometrical considerations, we can express $\sin \theta=\dfrac{D}{2R}$ where $2\theta$ is the angle of the deformed circular segment. Consequently, $\theta=\sin^{-1}({\kappa^*}/{2})$. The effective  stretch-ratio in this deformed state can then be written as $\lambda=\lambda_0\dfrac{2\theta R}{D}$, where $\lambda_0$ is the initial pre-stretch applied. Substituting this, we obtain:
\begin{equation}
\lambda= \dfrac{2\lambda_{0}}{\kappa^{*}}\sin^{-1}({\kappa^{*}}/{2}),
\label{kappa_lambda_sine_eqn}
\end{equation}
where $\kappa^{*} = \kappa D$.
\textcolor{black}{Expanding the inverse sine for small values of $\kappa^*$, and retaining terms to ${\cal O}(\kappa^{*3})$, we get $$\sin^{-1}(\kappa^*/2) \approx \kappa^*/2+\kappa^{*3}/48$$ and substituting this in Eq.~\ref{kappa_lambda_sine_eqn}, the curvature can be written as 
\begin{equation}
\kappa^*\approx 5 \sqrt{\lambda/\lambda_0-1}.
\label{eqn:curvature_lambda}
\end{equation}}

Next, we extend the uniaxial Gent relation to  equi-biaxial stretch condition. Assuming incompressibility of the material, the deformation gradient tensor ($\mathbf{F}$) for equi-biaxial extension can be written as a diagonal matrix with  elements: $\lambda, \lambda$, and $\dfrac{1}{\lambda^{2}}$ respectively. The left Cauchy-Green tensor,
\begin{equation}
\mathbf{B}=\mathbf{F}.\mathbf{F^{T}}.
\end{equation}
The Cauchy stress ($\mathbf{\sigma}_b$) for incompressible Gent model can be written as,
\begin{equation}
\mathbf{\sigma}_b=-p\mathbf{I}+G_m\mathbf{B}, 
\end{equation}
and so the in-plane stresses are,
\begin{equation}
\sigma_{11}=\sigma_{22}=G_m(\lambda^{2}-\dfrac{1}{\lambda^{4}}).
\end{equation}
Further, we derive the tension,  $T_b$, from the first Piola-Kirchhoff stress,
\begin{equation}
T_b = G_m h(1-\dfrac{1}{\lambda^{6}}).
\end{equation}

\begin{figure}[!t]
	\centering{
		\includegraphics[width=0.45\textwidth]{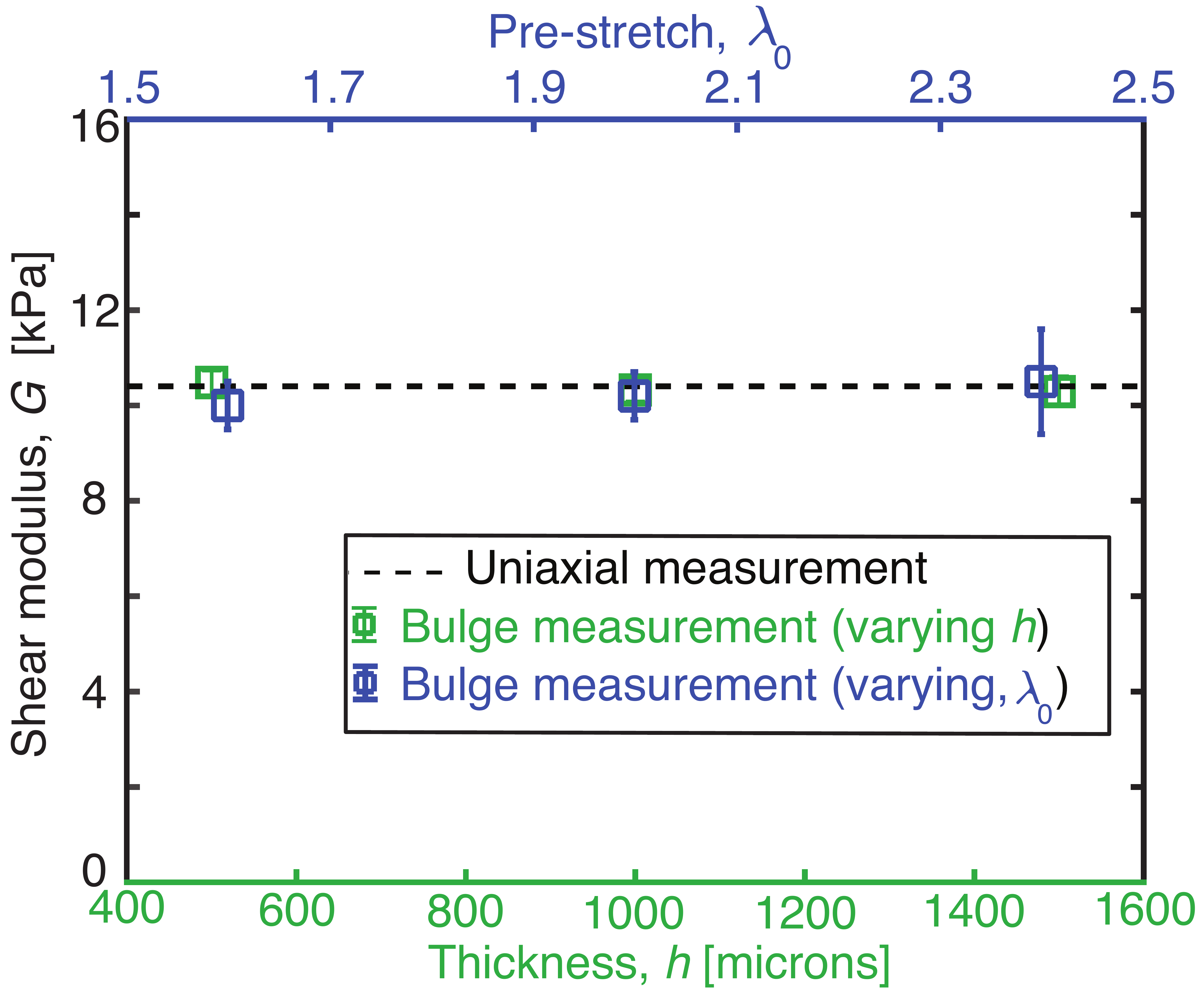}}
	\caption{Depicts the comparison between shear modulus estimates obtained using uniaxial and bulge tests. The horizontal dashed line corresponds to the uniaxial test estimate. The green squares correspond to cases with varying thickness, whereas the blue squares depict the cases with different pre-stetches. Error bars represent the standard deviation of the data. }
	\label{figS:Gvsh}
\end{figure}

The static equilibrium shape of a membrane under a pressure loading is approximated by the Young–Laplace equation \cite{waldman2017camber},
\begin{equation}
\kappa+ \dfrac{p}{T_b}=0
\end{equation}
\textcolor{black}{Using the relation from Eg. \ref{eqn:curvature_lambda} and }substituting the constitutive relations under biaxial loading conditions, we obtain an relation for the pressure as,
\begin{equation}
p \approx\dfrac{10G_m h}{D}(1-\dfrac{1}{\lambda^{6}})(\dfrac{\lambda}{\lambda_{0}}-1)^{1/2}.
\end{equation}

~\\

\section{Model predictions of Shear modulus}

Fig.~\ref{figS:Gvsh} shows a comparison of the shear modulus, $G$, predicted independently using uniaxial and circular bulge tests. The predicted shear modulus is unchanged for different values of thickness (green data points)  and pre-stretch (blue data points) tested.

\section{Supplemental Videos}

Supplemental video S1: Unwrinkling of membrane \href{https://www.dropbox.com/s/9wnxeswibgyxuch/video_fast.mp4?dl=0}{\tt[link]}

Supplemental video S2: Biaxial Stretching device operation \href{https://www.dropbox.com/s/8742nj1rmuzic3h/video_stretch.mp4?dl=0}{\tt[link]}
\\

\section{CAD files for Stretcher fabrication}
Part files for CAD assembly of biaxial stretcher. \href{https://www.dropbox.com/sh/mkkydbu33lrrhz2/AAALtpRnGAS4P8AXn5EXNjF3a?dl=0}{\tt [link]}

~\\

%\bibliographystyle{prsty_allauthors}
%\bibliography{aipsamp}

%merlin.mbs aipnum4-1.bst 2010-07-25 4.21a (PWD, AO, DPC) hacked
%Control: key (0)
%Control: author (8) initials jnrlst
%Control: editor formatted (1) identically to author
%Control: production of article title (0) allowed
%Control: page (1) range
%Control: year (1) truncated
%Control: production of eprint (0) enabled
\providecommand{\noopsort}[1]{}\providecommand{\singleletter}[1]{#1}%